\documentclass[aps,pra,epsfigure,twocolumn,showpacs]{revtex4}
\usepackage{amsmath}
\usepackage{amstext}
\usepackage{latexsym}
\usepackage{psfig}
\usepackage{graphicx}
\usepackage{amsfonts}
\usepackage{amssymb}

\newcommand{\ket}[1]{\left\vert#1\right\rangle}
\newcommand{\modul}[1]{\left\vert#1\right\vert}

\newcommand{\one}{\mbox{$1 \hspace{-1.0mm}  {\bf l}$}}

\newcommand{\pro}[2]{\left\vert#1\rangle\langle#2\right\vert}

\newcommand{\proj}[3]{\left\vert#1\rangle_{#2}\langle#3\right\vert}
\newcommand{\bra}[1]{\left\langle#1\right\vert}

\begin{document}

\title{Qubit state guidance without feedback}
\author{M. Paternostro and M. S. Kim}
\affiliation{School of Mathematics and Physics, The Queen's University, Belfast, BT7 1NN, UK}
\date{\today}

\begin{abstract}
We study a protocol for two-qubit state guidance that does not rely on feedback mechanisms. In our scheme, entanglement can be concentrated by arranging the interactions of the qubits with a continuous variable ancilla. By properly post-selecting the outcomes of repeated measurements of the state of the ancilla, the qubit state is driven to have a desired amount of purity and entanglement.
 We stress the primary role played by the first iterations of the protocol. Inefficiencies in the detection operations can be fully taken into account.
\end{abstract}
\pacs{03.67.-a, 03.67.Hk, 42.50.-p}
\maketitle

\section{Introduction}
\label{introduzione}

Quantum control is the ability to drive a quantum system toward desired states having disparate and arbitrary properties in terms of degree of purity and entanglement. The proposal and realization of schemes for the guidance of physical devices is central to the tasks of quantum information processing (QIP) as many of the protocols proposed so far may be formally interpreted in terms of control of an apparatus operated through classical or quantum potentials~\cite{nielsen}. However, it is commonly accepted that any external interference on a physical device opens channels to the interaction of the system itself with the uncontrollable environment, leading to decoherence~\cite{zurek}. Many different strategies have been suggested, in these years, to correct or avoid the effects of decoherence, ranging from quantum error-correcting codes~\cite{ecc} to quantum computing in decoherence-free subspaces~\cite{dfs} or dynamical-decoupling techniques~\cite{dd}. Another possible strategy is represented by the use of feedback to recursively modify the dynamics of a system.

A common feature of all these proposals is that they operate directly on the system to control. That is, {\it local} guidance is performed. However, it would be desirable to have a procedure allowing for a non-intrusive driving of a quantum device. In this case, the quantum control will be {\it remote} in the sense that the active part of a guidance protocol (including unitary and non-unitary operations) should be realized on an ancilla connected to the system we are interested in. This strategy is interesting, for example, for the purpose of creating a quantum channel to be used for later tasks of a processing protocol. The channel could be generated with minimal interventions of its constituents and its purity and entanglement could be regulated acting on a remote {\it knob}. In this paper, we address some recently proposed schemes for the purification of a quantum state and entanglement creation~\cite{nakazato,lidar} under the point of view of quantum remote control. We show how these schemes can be suitably modified in order to perform quantum guidance of a two-qubit state.

In details, Nakazato {\it et al.}~\cite{nakazato} have suggested a scheme for the probabilistic purification of the state of a quantum system (labeled B from now on). The scheme is based on the repeated measurements of the state of an ancillary system, here indicated as A, which interacts with B through some local Hamiltonian $\hat{H}_{int}$. System B is prepared in a state $\rho_{B}$ we want to purify while system A is initialized in $\ket{\chi}_{A}$. Let us assume that we are able to implement the projective measurement $\hat{P}_{A}=\proj{\chi}{A}{\chi}$. We apply the following program: A+B evolve through $\hat{H}_{int}$ for a time interval $\Delta{t}$ and state A is checked to be in $\ket{\chi}_{A}$. Tracing out the ancillary system, B evolves as 
\begin{equation}
\label{zero}
\begin{aligned}
\rho_{B}(\Delta{t})&=\mbox{Tr}_{A}(\hat{P}_{A}e^{-i\hat{H}\Delta{t}}\proj{\chi}{A}{\chi}\otimes\rho_{B}(0)e^{i\hat{H}\Delta{t}}\hat{P}_{A})\\
&=\hat{O}_{\chi}\rho_{B}(0)\hat{O}^{\dagger}_{\chi},
\end{aligned}
\end{equation}
 where $\hat{O}_{\chi}=_{A}\!\!\bra{\chi}{e^{-i\hat{H}\Delta{t}}}\ket{\chi}_{A}$. The observation by Nakazato {\it et al.} is that, if in the spectral decomposition of $\hat{O}_{\chi}$ there is a single term dominating over the rest of the spectrum, then system B is progressively purified by the repeated application of this evolution-detection protocol. The scheme is probabilistic with a probability of success ({\it i.e.} the probability that $\rho_{B}$ is driven toward a pure state $\ket{v}_{B}$) that critically depends on $_{B}\!\bra{v}\rho_{B}(0)\ket{v}_{B}$~\cite{nakazato}. It is worth stressing that, due to the non-unitary nature of the projector $\hat{P}_{A}$, it is impossible to describe the effective dynamics of subsystem B alone using the operator-sum representation~\cite{nielsen}. 

More recently, it has been recognized~\cite{lidar} that this same scheme may be generalized to the cases in which the system to purify is intrinsically multipartite (subparties $B_{1},B_{2},..,B_{m}$). In this case, the repeated measurements operated on the ancilla A project system B onto a pure entangled state of the subsystems $B_{i}\,(i=1,..,m)$. 

In this work, we assess the protocol for the generation of entangled pure states of a composite system B under the point of view of quantum guidance of a system. We heuristically develop the intuition (implicitly underlying ref.~\cite{nakazato}) that the effectiveness of the purification scheme through repeated measurements ({\it Zeno-like measurements}, in the language of Nakazato {\it et al.}~\cite{nakazato}) may be ascribed to some kind of quantum control procedure. We suggest a scheme that is nearer to experimental state of the art. In our analysis, indeed, we get rid of the unrealistic assumption of measurements projecting the ancilla onto the {\it bona fide} state $\ket{\phi}_{A}$ (a condition that is challenging in many practical cases). Furthermore, no feedback is required, in our procedure, but just the post-selection of favourable events and the proper control of the interaction interval. We show that, considering the ancillary system A to be embodied by a bipartite continuous variable (CV) system, a wide range of possibilities for quantum state guidance is offered. To the best of our knowledge, this point was never addressed in this context. Just one-qubit systems were used to embody the ancilla~\cite{nakazato,lidar}. 


\section{The system}
\label{sistema}

We set the scenario in order to show that a two-mode CV {\it mediator}, in a proper initial entangled state, is able to drive two qubits, labelled $1$ and $2$, toward a pure state, distilling entanglement at the same time. 

\begin{figure}[ht]
\centerline{\psfig{figure=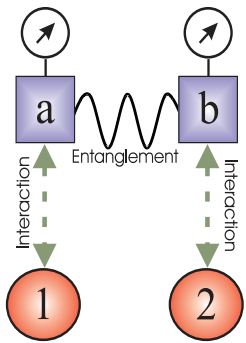,width=3.2cm,height=4.5cm}}
\caption{Scheme of the protocol for quantum control without feedback. The entangled two-mode ($a$ and $b$) CV system interacts with the (initially separable) qubits $1$ and $2$ respectively. After the interaction, the state of $a$ and $b$ is detected and the events are post-selected. The conditioned qubit state obtained after $n$ iterations of this program is progressively purified and its entanglement is distilled as $n$ gets larger.}
\label{principio}
\end{figure}

We consider a system of two remote qubits, $1$ and $2$, each defined in Hilbert spaces spanned by the basis $\{\ket{g},\ket{e}\}_{i}$ $(i=1,2)$. The qubits are initially prepared in a separable state and are mutually independent (there is no direct interaction between the qubits all along the quantum control protocol). The qubits constitute the subparties of system B whose state we want to control. Each qubit interacts with a CV system ($1$ interacts with $a$ and $2$ with $b$) which is in an entangled state. We consider local interactions $\hat{H}_{int}=\hat{H}_{a1}+\hat{H}_{b2}$ between each qubit and the corresponding CV mode. A sketch of the scheme we consider is shown in Fig.~\ref{principio}. After the subsystems mutually interact via $\hat{H}_{int}$ for a time interval $\Delta\tau$, $a$ and $b$ are measured~\cite{commento1}. 

A good source of quantum correlated two-mode states of a CV system is a non-degenerate-parametric-amplifier generating a two-mode squeezed vacuum (TMSV) of its squeezing parameter $s$~\cite{knightsqueezed}. On the other hand, it has been recently shown that the fidelity of teleportation of a coherent state can be improved~\cite{cochrane} and the loophole-free non-locality test is possible~\cite{cerf} by conditioning the CV entangled resource (embodied in a TMSV) through linear optical elements and photodetection~\cite{opatrny}. The specific form of the state considered in~\cite{cochrane,cerf} is the {\it photon-subtracted state} (PSS)
\begin{equation}
\label{PS}
\ket{\chi}_{ab}={\cal N}\sum^{\infty}_{l=0}\lambda(l)\ket{l,l}_{ab},
\end{equation} 
where $\lambda(l)=(\tanh{s})^{l}(l+1)$, $\ket{l}$ is a photon-number state and ${\cal N}$ is a normalization. The probabilistic generation of Eq.~(\ref{PS}) using a TMSV, high-transmittivity beam-splitters and photodetectors is described in ref.~\cite{cochrane}.

As in the teleportation protocol, the ancilla A$=(a,b)$ represents a resource. Its role, in this paper, is the catalysis of the purification process and the distillation of entanglement between the qubits. Under this point of view, we have to look for the best choice for system A. Even though the entire analysis and the results we report in this work can be generalized to any form of a two-mode quantum correlated state, we have checked that the choice of a PSS offers a good result in terms of entanglement and purity for the qubit state. This is essentially due to the fact that the conditioned way in which a PSS is generated from a TMSV effectively results in an entanglement distillation procedure. This can be seen noticing that in a TMSV, the photon-number distribution $P_{TMSV}(l)$ ({\it i.e.} the probability amplitude for the two-mode photon-number state $\ket{l,l}_{ab}$) is $P_{TMSV}(l)=(\tanh{s})^{l}/\cosh{s}$, which is rapidly decreasing function of $l$. In a PSS, on the other hand, the presence of the $(l+1)$ factor shifts $P_{PSS}(l)$ toward higher photon-number states. This results, in turns, in a higher von Neumann entropy~\cite{nielsen} for the PSS than for the TMSV, at any value of the squeezing parameter $s$~\cite{cochrane}. The main difficulty related to $\ket{\chi}_{ab}$ is the small probability with which such a state is generated~\cite{cochrane}. However, this is certainly not the central point of our investigation. We will consider the generation of the PSS as an off-line resource and, for the sake of generality, we do not specify the physical system chosen to embody the qubits. 


\section{The ideal protocol} 
\label{ideale}

\begin{figure}[ht]
\centerline{\psfig{figure=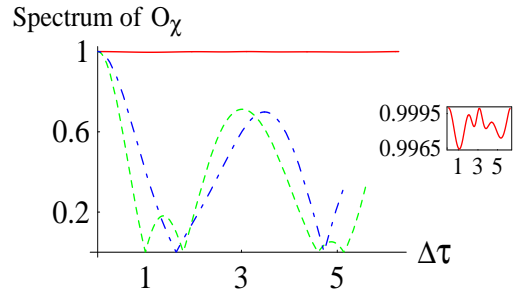,width=7.5cm,height=5.0cm}}
\caption{Eigenspectum of the post-measurement operator $\hat{O}_{\chi}$ in Eq.~(\ref{naka}) for $\ket{\chi}_{ab}$ a photon-subtracted state of $s=0.3$. The eigenvalues $e_{+}$ (solid line), $e_{-}$ (dashed) and $\modul{e_{0}}$ (dot-dashed) are presented as functions of the (dimensionless) interaction time $\Delta\tau$. Inset: the behavior of $e_{+}$ against $\Delta\tau$ is plotted in a shorter vertical scale to show the amplitude of its oscillations.}
\label{nakaz}
\end{figure}

Having identified the form of the state for our ancilla, we can apply the idealized protocol for purification. We have to calculate the structure of the operator $\hat{O}_{\chi}$ analogous to the one in Eq.~(\ref{zero}). In our case, it is 
\begin{equation}
\label{naka}
\begin{split}
\hat{O}_{\chi}&=O_{11}\pro{gg}{gg}+O_{44}\pro{ee}{ee}+O_{12}(\pro{eg}{eg}+\pro{ge}{ge})\\&-O_{14}(\pro{ee}{gg}+\pro{gg}{ee}),
\end{split}
\end{equation}
where $O_{11}={\cal N}^2\sum^{\infty}_{l=0}\lambda^2(l){\cal Q}^2_{l}$, $O_{44}={\cal N}^2\sum^{\infty}_{l=0}\lambda^{2}(l){\cal Q}^2_{l+1}$, $O_{12}={\cal N}^2\sum^{\infty}_{l=0}\lambda^{2}(l){\cal Q}_{l}{\cal Q}_{l+1}$ and $O_{14}=-{\cal N}^2\sum^{\infty}_{l=0}\lambda(l)\lambda(l+1)\sqrt{1-{\cal Q}^2_{l+1}}$. Here, ${\cal Q}_{l}=\cos(\Delta\tau\sqrt{l})$. To check the effectiveness of the purification procedure, we have to look for the eigendecomposition of $\hat{O}_{\chi}$. We find that, together with the double-degenerate eigenvalue $e_{0}=O_{12}$, the spectrum of $\hat{O}_{\chi}$ includes the eigenvalues $e_{\pm}=[(O_{11}-O_{44})\pm\sqrt{(O_{11}-O_{44})^2+4O^2_{14}}]/2O_{14}$ which correspond to the eigenoperators $\ket{\pm}_{12}\bra{\pm}$ respectively. Here, $\ket{\pm}_{12}=N_{\pm}[(e_{\pm}-O_{44})\ket{gg}+O_{14}\ket{ee}]_{12}$. The eigenspectrum is shown in Fig.~\ref{nakaz} where it is evident that $\modul{e_{+}}$ is much larger than $\modul{{e_{0}}}$ and $\modul{{e_{-}}}$ at any $\Delta\tau>{0}$. In particular, at $\Delta\tau\simeq{4.5}$, we find $e_{+}\simeq{0}.998\gg{e}_{-}=\modul{e_{0}}\simeq{0.028}$. This means that by repeating the protocol $n$ times and post-selecting the favourable events in which the subsystem A=$(a,b)$ is detected in $\ket{\chi}_{ab}$, the joint state of the qubits $1$ and $2$ is projected onto a state with an increasing degree of purity and entanglement (even if, given that $\modul{e_{+}}<1$, the state can not be maximally entangled). In order to quantify the degree of entanglement, we use the entanglement measure  based on {\it negativity of partial transposition} (NPT)~\cite{npt}. NPT is a necessary and sufficient condition
for entanglement of any bipartite qubit state~\cite{npt}. The corresponding degree of entanglement is defined as
${\cal E}_{NPT}=\max\{0,-2\epsilon^{-}\}$ with $\epsilon^{-}$ the negative
eigenvalue of the partial
transposition of the qubit density matrix. With this tool, we find that for qubits initially prepared in their ground states, after just one iteration, the degree of entanglement for the resulting two-qubit state is ${0.798}$, which is improved to $0.858$ after $n=5$ iterations~\cite{commento2}. 

However, in this protocol, a way to reliably perform the von Neumann measurements onto the prepared state of the ancilla is required. Checking exactly whether or not the ancilla is in its state $\ket{\chi}$ is, in general, a very hard task under a practical point of view. Even if in the case of qubits embodying the ancillary system A (that is basically the case considered in refs.~\cite{nakazato,lidar,compagno}) one may envisage some strategy to perform the desired projections, this seems not to be the case when a CV system is used. This should strongly affect the performances of the entire procedure. Nevertheless, a way to bypass the problem represented by the measurement step can be found, as it is argued in the next Section.

\section{Guidance without feedback} 
\label{nostro}

The suggestions in refs.~\cite{nakazato,lidar} to purify a qubit state and to distill entanglement are certainly intriguing due to the possibility of conditionally modifying the state of a system without directly interfering with it in a non-unitary way. Here, we look for a way to retain the basic idea behind these schemes, {\it i.e.} the modifications of the qubit state by interventions on an ancillary system. However, we abandon the idealized projective measurements.

The aim of this Section is to show that, with suitable modifications of the original protocol, the problem represented by the difficult von Neumann projections may be bypassed without affecting the effectiveness of the procedure. We consider a more realistic detection model that can be generalized to include arbitrary detection inefficiencies. Our protocol allows us to track the evolution of the two-qubit state showing that true quantum control can be performed with the qubits being actively driven toward a desired state. 

While the system we consider remains essentially the same as described in the previous Section, here we introduce a {\it near to reality} detection step. We borrow the language from the quantum optics language and model the detection step by Geiger-like on/off detectors that discriminate the presence or absence of excitations in the CV states, irrespective of the excitation number. The quantum efficiency of each detector (assumed to be the same) is $\eta$. An inefficient detector is formally described by the positive-operator-valued-measure (POVM)~\cite{nielsen} 
\begin{equation}
\label{povm}
\hat{\Pi}^{i}_{nc}({\eta})=\sum_{n=0}^{\infty}(1-\eta)^{n}\proj{n}{i}{n},\hskip0.3cm
\hat{\Pi}^{i}_{c}({\eta})=\one-\hat{\Pi}^{i}_{nc}({\eta}),
\end{equation}   
where $i=a,b$. The effect of detection inefficiencies on the quantum control protocol will be fully taken into account later on. For a while, we consider perfect detectors of $\eta=1$, so that $\hat{\Pi}^{i}_{nc}(1)\equiv\hat{\Pi}^{i}_{nc}=\ket{0}_{i}\!\bra{0}\,(i=a,b)$. Furthermore, at each step of the control procedure, the qubits will interact with a fresh quantum correlated state of modes $a$ and $b$. That is, there is no direct feedback of the ancilla state (another hard task in the original protocol \cite{nakazato,lidar}) but the qubits interact repeatedly with the fresh ancilla.

We consider the evolution of the qubit system resulting from the following program: the quantum correlated state of the CV system is prepared and the qubit register is initialized as $\rho_{12}(0)$. The subsystems interact through the joint interaction $\hat{H}_{int}$ that gives rise to the direct product of local unitaries $\hat{U}_{12ab}(t)=\hat{U}_{1a}(t)\otimes\hat{U}_{2b}(t)$, where $\hat{U}_{ij}=e^{-i\hat{H}_{ij}t}$ ($i=a,b$ and $j=1,2$). For the sake of definiteness, we assume the Hamiltonian model $\hat{H}_{ij}$ to be of the resonant Jaynes-Cummings form~\cite{shore}. This is a natural model that turns out to be valid in many physical situations in which coherent exchange of excitations between spin-like particles and bosonic modes are involved~\cite{noi,wilsonrae} (however, our study can be extended to any other entangling qubit-mode interaction). 

After the evolution, the state of the CV system is detected. The event in which both the detectors click is retained and the entire process is repeated $n$ times. A single step of this program changes the qubit state as 
\begin{equation}
\label{qubitevolution}
\begin{aligned}
\rho_{12}(n+1,\Delta\tau)&=\mbox{Tr}_{ab}[(\hat{\Pi}^{a}_{c}\otimes\hat{\Pi}^{b}_{c})\hat{U}_{12ab}(\Delta\tau)\rho_{12}(n)\\&
\otimes\ket{\chi}_{ab}\!\bra{\chi}\hat{U}^{\dagger}_{12ab}(\Delta\tau)(\hat{\Pi}^{a}_{c}\otimes\hat{\Pi}^{b}_{c})]
\end{aligned}
\end{equation}
with $n\ge{0}$. Here, we have explicitly indicated that the density matrix depends on the step $n$ performed and the (dimensionless) interaction interval $\Delta\tau$ which is supposed to be equal for the two qubit-mode subsystems. Whenever a {\it no-signal} event interrupts a sequence (corresponding to either one or both the detectors failing to click), the two-qubit state is discarded and the process is re-started.

It is not difficult to evaluate explicitly the effect of the operator $(\hat{\Pi}^{a}_{c}\otimes\hat{\Pi}^{b}_{c})\hat{U}_{12ab}(t)$ after the first step of the procedure. For definiteness, we assume $\rho_{12}(0)=\ket{gg}_{12}\bra{gg}$~\cite{commento}. Using the resolution of the identity operator $\one_{i}=\sum^{\infty}_{l=0}\ket{l}_{i}\!\bra{l}$ ($i=a,b$) and the photon-number basis to compute the trace in Eq.~(\ref{qubitevolution}), we get

\begin{equation}
\label{esplicito1}
\begin{aligned}
\rho_{12}(1,\Delta\tau)&={\cal A}_{gggg}\pro{gg}{gg}+{\cal B}_{gggg}(\pro{ge}{ge}+\pro{eg}{eg})\\&+{\cal F}_{gggg}\pro{ee}{ee}+{\cal G}_{gggg}(\pro{gg}{ee}+\pro{ee}{gg}),
\end{aligned}
\end{equation}
where ${\cal A}_{gggg}={\cal N}^{2}\sum^{\infty}_{l=1}\lambda^{2}(l){\cal Q}^{4}_{l}$, ${\cal F}_{gggg}={\cal N}^{2}\sum^{\infty}_{l=2}\lambda^{2}(l)(1-{\cal Q}^{2}_{l})^2$, ${\cal B}_{gggg}={\cal N}^{2}\sum^{\infty}_{l=2}\lambda^{2}(l){\cal Q}^{2}_{l}(1-{\cal Q}^{2}_{l})$, ${\cal G}_{gggg}=-{\cal N}^{2}\sum^{\infty}_{l=1}\lambda(l)\lambda(l+1){\cal Q}^{2}_{l}(1-{\cal Q}^{2}_{l+1})$. In what follows, ${\cal A}$ stands for the coefficient relative to the $_{12}\!\bra{gg}\rho_{12}(n,t)\ket{gg}_{12}$ density matrix elements (the same holds, {\it mutatis mutandis}, for the other coefficients) while the subscripts will label the element of the initial density matrix we are considering. 
It is important to stress the correspondence between the form of $\rho_{12}(1,\Delta\tau)$ and that of the effective post-measurement operator $\hat{O}_{\chi}$ in Eq.~(\ref{naka}). Indeed, the application of Eq.~(\ref{naka}) to the initial state $\rho_{12}(0)=\ket{gg}_{12}\bra{gg}$ will result in a density matrix having exactly the same form as Eq.~(\ref{esplicito1}). This is an effect due to the particular symmetries in the quantum-correlated state of the ancilla and it assures that our protocol reproduces the right form of the density matrix. Our task is to show that the degree of entanglement and purity for the qubits progressively increase as the protocol is repeated.

It is straightforward to prove that the form of a state as Eq.~(\ref{esplicito1}) is invariant for successive applications of the dynamics described by Eq.~(\ref{qubitevolution}). The density operator is trapped in a form that deserves some comments. The absence of the coherences $_{12}\!\bra{eg}\rho_{12}(1,\Delta\tau)\ket{ge}_{12}$ and its Hermitian conjugate and the fact that the populations $_{12}\!\bra{ge}\rho_{12}(1,\Delta\tau)\ket{ge}_{12}$ and $_{12}\!\bra{eg}\rho_{12}(1,\Delta\tau)\ket{eg}_{12}$ are equal suggest that, on the basis of the maximally entangled Bell states $\{\ket{\phi_{+}},\ket{\phi_{-}},\ket{\psi_{+}},\ket{\psi_{-}}\}_{12}$~\cite{nielsen}, Eq.~(\ref{esplicito1}) can be written as the mixture 
\begin{equation}
\begin{aligned}
\rho_{12}(1,\Delta\tau)&=\alpha(\ket{\psi_{+}}_{12}\!\bra{\psi_{+}}+\ket{\psi_{-}}_{12}\!\bra{\psi_{-}})\\
&+\beta\ket{\phi_{+}}_{12}\!\bra{\phi_{+}}+\gamma\ket{\phi_{-}}_{12}\bra{\phi_{-}}.
\end{aligned}
\end{equation}
Here, $\ket{\phi_{-}}_{12}=(1/\sqrt{2})(\ket{gg}-\ket{ee})_{12},\,\ket{\phi_{+}}_{12}=\hat{\sigma}^{z}_{1}\ket{\phi_{-}}_{12},\,\ket{\psi_{-}}_{12}=\hat{\sigma}^{x}_{1}\ket{\phi_{-}}_{12}$ and $\ket{\psi_{+}}_{12}=i\hat{\sigma}^{y}_{1}\ket{\phi_{-}}_{12}$~\cite{nielsen}. The purity and entanglement of the final state depend on the relative weight of the coefficients $\alpha,\beta$ and $\gamma$.

\begin{figure}[ht]
\centerline{\psfig{figure=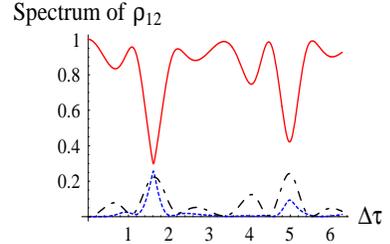,width=5.0cm,height=3.5cm}}
\caption{Eigenspectrum of the density matrix for the two-qubit state after a single application of the quantum control protocol against the rescaled time interval $\Delta\tau$. The eigenvalue $\beta_{+}$ (solid line) dominates over the rest of the eigenspectrum. There is a double-degenerate eigenvalue, which is why there are just three curves, in this plot. As in Fig.~\ref{nakaz}, here the squeezing parameter in the photon-subtracted state is $s=0.3$.}
\label{spettro}
\end{figure}

The first indications about the state that is expected and its corresponding degrees of purity and entanglement are given by the eigenspectrum of $\rho_{12}(1,\Delta\tau)$. As a consequence of the analysis above, it is not surprising that the eigenspectrum presents exactly the same characteristics pointed out about $\hat{O}_{\chi}$ (see  the discussion after Eq.~(\ref{naka}), in Section~\ref{ideale}), {\it i.e.} the existence on an eigenvalue that dominates over all the others. We indicate this eigenvalue as $\beta_{+}=[({\cal A}_{gggg}-{\cal F}_{gggg})+\sqrt{({\cal A}_{gggg}-{\cal F}_{gggg})^2+4{\cal G}^2_{gggg}}]/2{\cal G}_{gggg}$ and, in Fig.~\ref{spettro}, we plot it (with the rest of the spectrum) against $\Delta\tau$. The eigenvector corresponding to $\beta_{+}$ turns out to be the (unnormalized) entangled state $(\beta_{+}-{\cal F}_{gggg})\ket{gg}_{12}+{\cal G}_{gggg}\ket{ee}_{12}$, which has a large overlap with $\ket{\phi_{-}}_{12}$, in perfect correspondence with what is obtained using the results described in Section~\ref{ideale}.

We go on with the procedure summarized by Eq.~(\ref{qubitevolution}) and look for the statistical properties of the resulting qubit state. After the first iteration, each non-zero density matrix element becomes more and more involved. However, it is possible to find a recurrence pattern that expresses the (unnormalized) matrix elements after $n+1$ iterations ($n\ge{1}$) as a function of the matrix elements after the first and the $n$-th iterations as
\begin{equation}
\label{esplicitoaltri}
\begin{aligned}
{\cal M}(n+1)&={\cal N}_{n}[{\cal A}(n){\cal M}_{gggg}+{\cal B}(n)({\cal M}_{egeg}+{\cal M}_{gege})\\&+{\cal F}(n){\cal M}_{eeee}+{\cal G}(n)({\cal M}_{eegg}+{\cal M}_{ggee})],
\end{aligned}
\end{equation}
where ${\cal N}_{n}=[{\cal A}(n)+2{\cal B}(n)+{\cal F}(n)]^{-1}$ is the normalization of the density matrix after $n$ iterations and ${\cal M}={\cal A},{\cal B},{\cal F},{\cal G}$. Obviously, the matrix elements at $n=1$ depend on the choice of the initial density matrix and coincide with the coefficient given in Eq.~(\ref{esplicito1}) for $\rho_{12}(0)=\ket{gg}_{12}\!\bra{gg}$.

\begin{figure}[ht]
{\bf (a)}\hskip4cm{\bf (b)}
\centerline{\psfig{figure=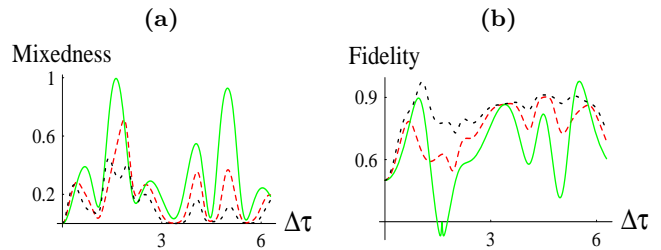,width=9.0cm,height=3.0cm}}
\caption{{\bf (a)}: Mixedness of the two-qubit state, as a function of the rescaled time $\Delta\tau$, after one (solid line), two (dashed) and three iterations (dotted) of the guidance protocol. {\bf (b)}: The corresponding fidelity between the cases considered in panel {\bf (a)} and the Bell state $\ket{\phi_{-}}_{12}$. In both the panels, we considered a squeezing parameter of the photon-subtracted state $s=0.3$.}
\label{purezzeoverlap}
\end{figure}

In order to characterize the state $\rho_{12}(n,\Delta\tau)$ we consider the degree of mixedness as measured by the linearized entropy ${S}_{L}(n)=(4/3)[1-\mbox{Tr}(\rho^{2}_{12}(n,\Delta\tau))]$~\cite{noi}. This quantity is $0$ for pure states and $1$ for completely mixed ones. Moreover, to get an immediate picture of how near the state we get is to the $\ket{\phi_{-}}_{12}$ Bell state, we consider the fidelity $F=_{12}\!\bra{\phi_{-}}\rho_{12}(n,\Delta\tau)\ket{\phi_{-}}_{12}$. These figures of merit have been plotted in Figs.~\ref{purezzeoverlap} {\bf (a)} and {\bf (b)} respectively, against the interaction interval $\Delta\tau$. Due to computational limitations, in these plots we report the results obtained for $n=1,2,3$. The results after further iterations can be evaluated numerically, pointwise with respect to the interaction interval.

Fig.~\ref{purezzeoverlap} {\bf (a)} shows that, in accordance with the results in Figs.~\ref{nakaz} and \ref{spettro}, the time interval $\Delta\tau\simeq{4.5}$ is interesting for purification tasks. When the interactions between subsystems A and B last for such a length of time, the repeated applications of our protocol lead to the reduction of the mixedness and the increase of the overlap with $\ket{\phi_{-}}_{12}$. The largest reduction in the mixedness value is obtained in the early iterations, when $S_{L}$ changes by nearly one order of magnitude from $S_{L}\simeq{0.09}$ to $S_{L}\simeq{0.01}$. The successive reductions are smaller but still considerable, up to the step $n=9$, which is the last we have considered in our calculations. 

It appears that different tasks may be achieved with our quantum control protocol simply choosing a different value of $\Delta\tau$. For example, it is possible to choose $\Delta\tau\simeq{3.3}$, corresponding to the situation in which successive applications of the guidance protocol increase the purity of the qubit state leaving its overlap with $\ket{\phi_{-}}_{12}$ unchanged. This realizes the purification of the qubit state for a given amount of entanglement shared between the qubits. This semi-quantitative analysis should have shown that genuine quantum control without any real feedback (but just post-selection) is performed, in this way. Other possibilities are offered by a {\it dynamic adjustment} of the interaction intervals, where the values of $\Delta\tau$ change at each iteration, in order to guide the state of the qubits toward a desired target state. However, it is obvious that such a procedure (representing a true feedback scheme) requires a much finer control on the dynamics of the setup. 

\begin{figure}[ht]
\centerline{\psfig{figure=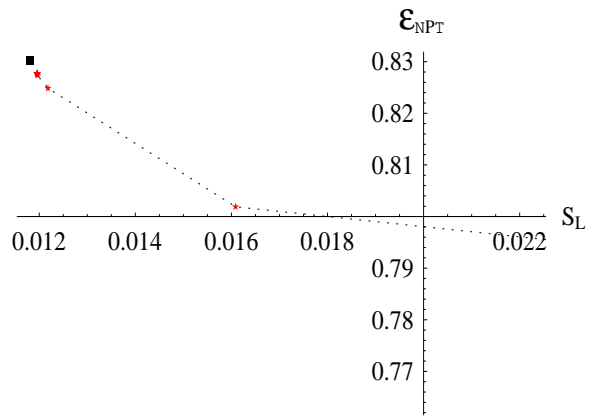,width=8.0cm,height=5.5cm}}
\caption{Comparison between the state after the ideal procedure (full square) and those obtained for $n=1\rightarrow{n=4}$ in our effective guidance protocol (stars). Here, $\Delta\tau=4.5$ and the squeezing parameter of the ancillary state is $s=0.3$. The evolution of the state described by the star-shaped symbol is from the right to the left of the horizontal axis.}
\label{comparo}
\end{figure}

The last step required to complete the analysis of the performances of our control protocol is a more direct comparison with the results achieved using the idealized procedure outlined in Section~\ref{ideale}. This may be done by contrasting the purity and entanglement of the qubit systems. As shown in Fig.~\ref{comparo}. The evolution of $\rho_{12}(n,\Delta\tau)$, in the purity-entanglement plane, is shown (from right to left) by the star-shaped symbols for up to $n=4$ iterations. We have taken $\Delta\tau=4.5$, in this plot. The black square is what is obtained from the idealized protocol. As we have pointed out, the larger reduction of the mixedness is accomplished after the early iterations. By reiterating our protocol, the improvement in the degree of purity (as well as the increases of the degree of entanglement) becomes smaller even if the purification of the state and its entanglement concentration are still effective. The series of points we obtain on the purity-entanglement plane slowly approaches the state produced by the ideal scheme in Section~\ref{ideale}.

\section{Non-idealities in the protocol}

\begin{figure}[ht]
\centerline{\psfig{figure=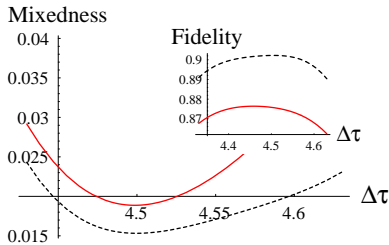,width=6.5cm,height=4.0cm}}
\caption{Comparison between the mixedness of the two-qubit state after two iterations of the quantum guidance protocol, as a function of the rescaled interaction time $\Delta{\tau}$, for perfect ($\eta=1$) Geiger-like detectors (dashed line) and for $\eta=0.7$ (solid line). Inset: comparison between the overlap of the two-qubit state with the $\ket{\phi_{-}}_{12}$ Bell state against $\Delta\tau$.}
\label{inefficienze}
\end{figure}

The effect of the detection inefficiencies can be fully taken into account. It turns out that the use of non-ideal detectors is formally equivalent to the replacement $\lambda(l)\rightarrow[1-(1-\eta)^{l}]\lambda(l)$ in Eq.~(\ref{PS}) followed by the explicit recalculation of Eq.~(\ref{qubitevolution}). Following these lines, we find that, for efficiency of detection as small as $\eta=0.7$, the mixedness of the qubit state is about $20\%$ larger than the case of perfect detectors, as it is shown in Fig.~\ref{inefficienze}. Even if Fig.~\ref{inefficienze} presents these results just for a rescaled time around $\Delta\tau\simeq{4.5}$, the same is true all along the relevant interaction times we have examined and for up to $n=9$ iterations of the control protocol. The same qualitative considerations hold for the overlap of the qubit state with the $\ket{\phi_{-}}_{12}$ Bell state (inset of Fig.~\ref{inefficienze}). Even if it is raised by the non-ideality of the photodetectors, the mixedness of the qubit state is, nevertheless, reduced at each further application of our procedure. It is thus clear that realistic inefficiencies at the detection step do not affect dramatically the guidance of the qubit state.

We can continue the analysis that non-ideality has in the performance of this protocol addressing the problem of {\it no-signaling} at the detection step. As pointed out earlier, the original purification scheme~\cite{nakazato} discards negative events in which the ancilla system is found in a state different from the bona fide one. In our protocol, a positive event is a coincidence at the detection stage. Even if we have demonstrated that a chain of detection coincidences is the desirable concatenation of events, the probability of such a result is certainly quite small. Our question is about the effect of a missing coincidence with respect to the purity of the qubit state. To answer this question, we have simulated a sequence of two applications of the procedure. The first application is {\it positive} (in the sense that a coincidence of detection is found) while the second one is {\it negative} (no signal at the detectors, which is formally described by the action of the operator $\otimes^{2}_{i=1}\hat{\Pi}^{i}_{nc}(\eta)$). It is straightforward to derive an explicit expression for the resulting density matrix and to evaluate the relative eigenspectrum. We find that the effectiveness of the purification process is not critically affected, even if the overlap with $\ket{\phi_{-}}_{12}$ is smaller than in the case of an ideal sequence of events. We have checked what happens to the qubit density matrix when the sequence of positive and negative events is inverted. In this case, the entanglement distillation process is no more effective, up to the second iteration. There is a very small increase in the overlap between the resulting qubit state and the target state $\ket{\phi_{-}}_{12}$. 

Another evidence of the importance of the first iteration can be found looking for the asymptotic structure of the density matrix elements. A numerical analysis shows that, within a reasonable degree of accuracy, the following approximations hold ${\cal F}(n+1)\simeq {\cal N}_{n}[{\cal F}(n){\cal F}_{eeee}+{\cal A}(n){\cal F}_{gggg}]$, ${\cal A}(n+1)\simeq {\cal N}_{n}[{\cal F}(n){\cal A}_{eeee}+{\cal A}(n){\cal A}_{gggg}], {\cal B}(n+1)\simeq{0}$ and ${\cal G}(n+1)\simeq {\cal N}_{n}[{\cal F}(n){\cal G}_{eeee}+{\cal A}(n){\cal G}_{gggg}]$. The final state of the qubit system at which the purification procedure ends, is obtained when the density matrix elements do not change any more after repeated application of the control protocol. More specifically, we are looking for $\rho_{12}(n,\Delta\tau)\simeq\rho_{12}(n+1,\Delta\tau)=\rho_{f}$, which is found to be
\begin{widetext}
\begin{equation}
\label{final}
\rho_{f}\simeq
\begin{pmatrix}
\frac{{\cal A}_{eeee}{\cal F}_{eeee}}{{\cal F}_{eeee}+{\cal A}_{eeee}-{\cal A}_{gggg}}&0&0&\frac{{\cal A}_{eeee}{\cal G}_{gggg}}{{\cal F}_{eeee}}+\frac{({\cal F}_{eeee}-{\cal A}_{gggg}){\cal G}_{eeee}}{{\cal F}_{eeee}}\\
0&0&0&0\\
0&0&0&0\\
\frac{{\cal A}_{eeee}{\cal G}_{gggg}}{{\cal F}_{eeee}}+\frac{({\cal F}_{eeee}-{\cal A}_{gggg}){\cal G}_{eeee}}{{\cal F}_{eeee}}&0&0&\frac{{\cal F}_{eeee}({\cal F}_{eeee}-{\cal A}_{gggg})}{{\cal F}_{eeee}+{\cal A}_{eeee}-{\cal A}_{gggg}}
\end{pmatrix}.
\end{equation}
\end{widetext}
This shows how, in pour case, the purity and entanglement of the final state entirely depend on the form of the density matrix after a single iteration of the purification protocol. The first step is the most important one and the control procedure has to be designed in such a way that the mixedness of $\rho_{f}$ is the smallest possible. It is worth noticing that, for our choice of quantum correlated CV state, this approximated asymptotic state leads to the values of the characterizing benchmarks equal to $S_{L}(f,\Delta\tau=4.5)\simeq10^{-2}$, ${\cal E}_{NPT}(f,\Delta\tau=4.5)\simeq 0.94$ and a fidelity of overlap with $\ket{\phi_{-}}_{12}$ $\simeq{0.91}$.
 
\section{Remarks}

We have studied the purification and entanglement distillation of a two-qubit system through its interaction with a CV ancilla on which repeated measurements are operated. This attempt is motivated by the somehow natural advantages in using photonic systems as ancillae. A Longer distance between the qubits may be achieved and more reliable detectors are available, for these systems. Taking advantage from our previous studies about entanglement-transfer processes~\cite{noi}, we have explicitly considered the case of a bipartite entangled state of the ancilla to show that an efficient and fast entanglement generation is possible, in this scenario. Our investigation is gone further in addressing the possibility of true quantum guidance of the state of two qubits without real feedback. This is certainly an advantage as feedback procedure is in general difficult to be implemented.

Our protocol is able to contemplate those inefficiencies, in the necessary detection steps, which are faced in realistic implementations of quantum optics setups. We have shown that the effectiveness of the guidance protocol is reasonably robust against negative detection events and realistic detector inefficiencies.

\acknowledgments

This work has been supported by the UK EPSRC and the KRF (2003-070-C00024). MP acknowledges IRCEP for financial support.


\begin{thebibliography}{99}
\bibitem{commento1} Excluding the detection step, a similar setup has been previously considered in relation to the entanglement-transfer problem~\cite{noi}. In that case, however, the control over the system's performances (here represented by the post-selection of favourable detection events) was totally absent, the overall procedure being much more passive than here.
\bibitem{commento2} It is worth stressing again that, using Eq.~(\ref{PS}), the amount of entanglement of the two-qubit state after just a single iteration of the protocol is larger than the corresponding value obtained using a TMSV.
\bibitem{commento} This assumption is just a convenient choice that makes the results of the calculations more transparent. Any initial state with any degree of mixedness can be considered. We are simply providing the necessary ingredients to evaluate the evolution of any arbitrary initial state.
\end{thebibliography}
\end{document}